\documentclass[12pt]{iopart}

\usepackage{graphicx}
\usepackage{dcolumn}
\usepackage{bm}
\usepackage{float}
\usepackage{afterpage}
\usepackage[colorlinks=true,
            linkcolor=black,
            urlcolor=black,
            anchorcolor=black,
            citecolor=blue,]{hyperref}
\begin{document}

\title{Peak Effects Induced by Particle Irradiations in 2\emph H-NbSe$_2$}

\author{Wenjie Li${^1{^,{^*}}}$, Sunseng Pyon${^1}$, Akiyoshi Yagi${^1}$, Cheng Yu${^1}$, Ryosuke Sakagami${^1}$, Ataru Ichinose${^2}$, Satoru Okayasu${^3}$, and Tsuyoshi Tamegai${^1}$}

\address{${^1}$Department of Applied Physics,The University of Tokyo, 7-3-1 Hongo, Bunkyo-ku, Tokyo 113-8656, Japan}
\address{${^2}$Grid Innovation Research Laboratory, Central Research Institute of Electric Power Industry, 2-6-1 Nagasaka, Yokosuka, Kanagawa 240-0196, Japan}
\address{${^3}$Advanced Science Research Center, Japan Atomic Energy Agency, Tokai, Ibaraki 319-1195, Japan}
\address{* Corresponding author: wenjiecd@gmail.com}
\vspace{10pt}

\begin{abstract}
Various peak effects in 2\emph H-NbSe$_2$ single crystals induced by particle irradiations were studied. 3 MeV proton irradiation magnified the peak effect induced by order-disorder transition of vortices, where the peak field shifts from high fields to low fields with increasing irradiation dose. For the peak effect in NbSe${\rm {_2}}$ with splayed columnar defects (CDs), as the splayed angle increases, peak field gradually shifts from high fields to low fields. Numerical calculations have been conducted to investigate the mechanism of the peak effect. The calculated results exhibit excellent agreement with experimental observations. Analyses of field dependence of \emph J${\rm {_c}}$ reveal the formation of non-uniform \emph J${\rm {_c}}$ flow with increasing splayed angle, which plays a crucial role in inducing the self-field peak effect in superconductors with splayed CDs. In samples with symmetric splayed CDs with respect to the \emph c-axis generated by 800 MeV Xe and 320 MeV Au ions, coexistence of order-disorder transition-induced peak effect and self-field peak effect was observed. In the case of 320 MeV Au irradiated samples, when the splay angle is small, the two peak effects transform into a broad peak, which has similarity to the anomalous peak effect observed in iron-based superconductors. Interestingly, the broad anomalous peak effect is strongly suppressed when the external magnetic field is applied parallel to one of splayed CDs.
\end{abstract}

\section{\label{sec:level1}Introduction}

In addition to the efforts to enhance the critical temperature (\emph T${\rm {_c}}$) of superconductors, improvements of upper critical field (\emph H${\rm {_{c_2}}}$) and critical current density (\emph J${\rm {_c}}$) are also very important for the practical application of superconductors \cite{hosono2018recent}. The \emph J${\rm {_c}}$ in superconductors is determined by the onset of vortex motion by the action of Lorentz force, which causes energy dissipation \cite{abrikosov2004nobel,blatter1994vortices,kwok2016vortices}. Therefore, understanding the mechanism of vortex pinning in superconductors is important for enhancing \emph J${\rm {_c}}$. When the pinning force is a constant, \emph J${\rm {_c}}$ should monotonically decrease with increasing applied magnetic field \emph H. However, in some cases \emph J${\rm {_c}}$ increases with increasing \emph H, resulting in formation of a peak in \emph J${\rm {_c}}$-\emph H curves (peak effect). A well-known example is the sharp peak which is frequently observed near \emph H${\rm {_c{_2}}}$ in many type II superconductors. It was initially explained by Pippard with the consideration that the rigidity of the flux lattice falls to zero rapidly close to \emph H${\rm {_c{_2}}}$ than the order parameter, which results in an enhancement of \emph J${\rm {_c}}$ in a narrow region close to \emph H${\rm {_c{_2}}}$ \cite{pippard1969possible}. At present we know that this peak effect is originated from a vortex phase transition from an ordered phase to a disordered phase near \emph H${\rm {_c{_2}}}$ \cite{paltiel2000instabilities}. Still, much work has to be done on this order-disorder (O-DO) transition-induced peak effect to understand any factors that affect the peak field (\emph H${\rm {_p}}$).

Another peak effect which is located at low magnetic fields is frequently observed in some superconductors with CDs introduced by heavy-ion irradiations \cite{tamegai2012effects}. CDs introduced into superconductors have structural similarity to vortices, and hence are expected to effectively pin vortices. Extensive studies on cuprate and iron-based superconductors have demonstrated significant improvements in \emph J${\rm {_c}}$ after the successful introduction of CDs \cite{tamegai2012effects,civale1991vortex,civale1997vortex,krusin1996enhanced,pradhan1998influence,fuchs2007strongly,holzapfel1993angle,haberkorn2015enhancement,moore2009effect,prozorov2010magneto,fang2012high}. The mechanisms of vortex pinning by CDs in superconductors has been theoretically explained \cite{nelson1993boson,nelson1992boson}. The low-field peak effect is known to arise from the self-field effect \cite{tamegai2012effects}, where the magnetic flux lines in thin superconductors under perpendicular field lower than the self-field (\emph H${\rm {_s{_f}}}$) are curved \cite{conner1991self} and cannot be effectively pinned by straight CDs perpendicular to the superconducting plane. As \emph H increases, those curved magnetic flux lines are straightened and can be effectively pinned by CDs, which results in a monotonic increase in \emph J${\rm {_c}}$ up to \emph H${\rm {_s{_f}}}$.

The third peak effect was also discovered in superconductors with CDs. The story started with the fact that there are many intrinsic defects such as twin boundaries in unconventional superconductors including cuprate superconductors. Those intrinsic defects play important roles in determining \emph J${\rm {_c}}$ \cite{pradhan1998influence}. To distinguish the effect of twin boundaries from those of artificially introduced CDs, tilted CDs (at an angle \emph {$\theta$}${\rm {_C{_D}}}$ from the \emph c-axis) were intentionally created \cite{civale1991vortex}, which results in the observation of an anomalous peak effect located at $\sim$1/3\emph B$_\Phi$ (\emph B$_\Phi$ is a dose equivalent matching field related to the density of CDs. \emph B$_\Phi$ = 1 T corresponds to 5×10{${^1{^0}}$}/cm{${^2}$} CDs) when \emph H is applied parallel to CDs \cite{civale1991vortex}. In conventional low-temperature superconductors such as NbSe${\rm {_2}}$, the effects of intrinsic defects on pinning are negligible, which means that it is not necessary to intentionally introduce tilted CDs to distinguish them from intrinsic defects. However, when parallel CDs were introduced into NbSe${\rm {_2}}$ along the \emph c-axis and \emph H was applied parallel to CDs, no peak effects were observed \cite{li2021peak}. Such an observation makes the study of the anomalous peak effect in superconductors with tilted CDs complicated. In principle, pinning of vortices would be very similar in superconductors with parallel CDs and tilted CDs when \emph H is applied parallel to the CDs. However, this intuitive understanding is inconsistent with experimental observations \cite{li2021peak}.

In addition to parallel and tilted CDs, it was theoretically proposed that \emph J${\rm {_c}}$ can be further enhanced by introducing splayed CDs into superconductors \cite{hwa1993flux}. This proposal has been confirmed by experiments where the splayed CDs can enhance \emph J${\rm {_c}}$ to a greater extent than parallel or tilted CDs in both cuprate and iron-based superconductors \cite{krusin1996enhanced,park2018field}. Furthermore, an anomalous peak feature similar to that observed in superconductors with tilted CDs had also been observed in superconductors with splayed CDs \cite{park2018field}. As far as we know, the origin of the anomalous peak effect in superconductors with neither tilted CDs nor splayed CDs is not fully understood.

In this paper, we conducted various particle irradiation experiments with different ion species and energies into 2\emph H-NbSe${\rm {_2}}$ single crystals to study the mechanisms of peak effects. To be specific, the O-DO transition-induced peak effect was observed after 3 MeV proton irradiation and how this peak effect was affected by the irradiation dose was studied. Coexistence of O-DO transition-induced peak effect and the self-field induced peak effect was observed in NbSe${\rm {_2}}$ when splayed CDs are introduced by 800 MeV Xe and 320 MeV Au irradiations. Different features of self-field peak effects in superconductors with parallel CDs and splayed CDs are explained. When the splay angle of CDs generated by 320 MeV Au irradiation is small in NbSe${\rm {_2}}$, an anomalous peak effect was observed, while it turns into the self-field peak effect at large splay angles. In addition, a general feature of the anomalous peak effect is found, where the peak is strongly suppressed when \emph H is applied parallel to one of the splayed CDs.

\section{\label{sec:level1}Experimental Details}
2\emph H-NbSe${\rm {_2}}$ single crystals were synthesized by iodine vapor transport method, where Nb and Se elements were mixed in the stoichiometric ratio with a total charge of 2 g. As a transport agent, 5 mg/cm${\rm {^3}}$ of iodine were sealed together into a quartz tube with a length of 10 cm and an inner diameter of 16 mm. The entire quartz tube was placed in a tube furnace, with one end containing the starting materials placed in the high-temperature zone (800 °C), while the low-temperature zone was set at ~750 °C. The vapor transport process took about two weeks. Single crystals were obtained in the high-temperature zone.

\begin{figure}[H]
\centering
\includegraphics[scale=0.5]{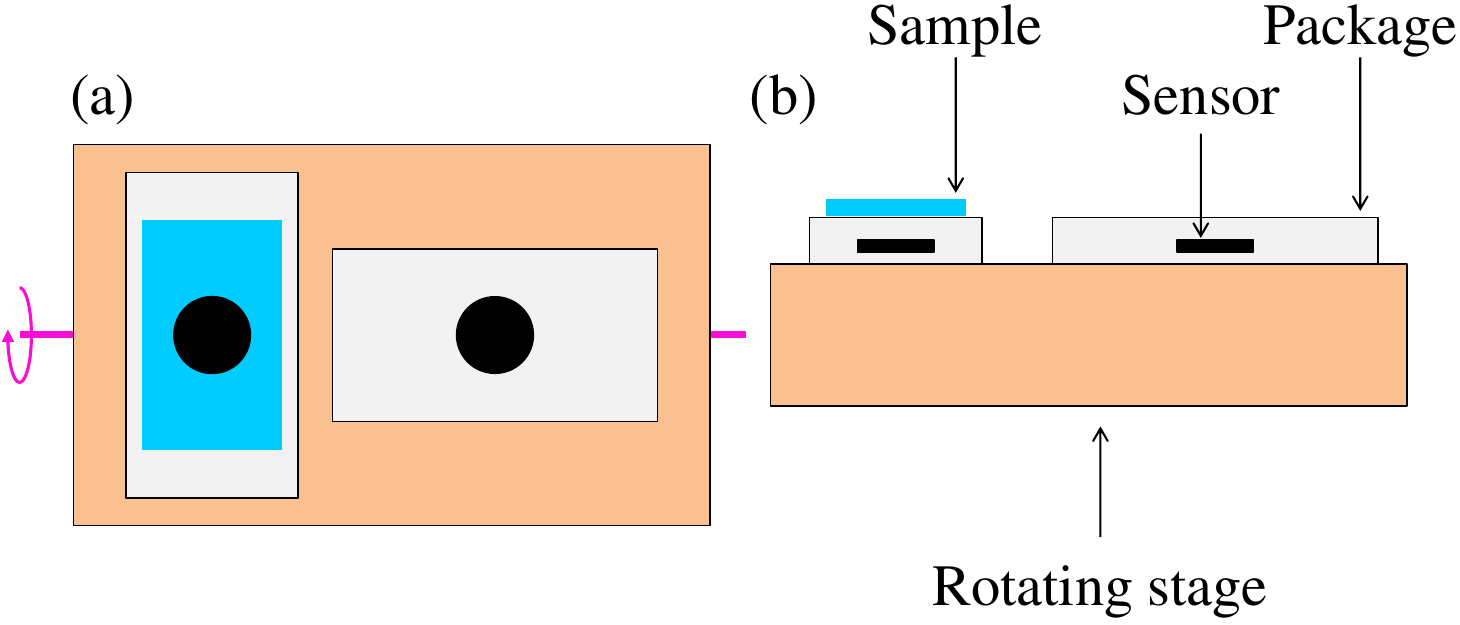}
\caption{\label{fig:1}Schematic diagrams of (a) the top view and (b) side view of the rotating stage, on which two miniature Hall probes are installed. One is used to measure the local magnetic induction of the sample, while the other is used to measure the external field. The rotating stage can be rotated from 0° to 180° around the axis perpendicular to the magnetic field.}
\end{figure}

3 MeV proton and 800 MeV Xe irradiations were conducted at NIRS-HIMAC in Chiba, Japan. 320 MeV Au irradiation experiments were conducted at JAEA in Ibaraki, Japan. To ensure that the CDs are introduced homogeneously throughout the whole sample, the samples were cleaved with a thickness of {${\sim}$}10 {${\mu}$}m. This thickness is much smaller than the projected range of energetic particles used in the present study in NbSe${\rm {_2}}$. The projected range was calculated by SRIM-2008 (the Stopping and Range of Ions in Matter-2008) \cite{ziegler1985stopping}. All irradiations were performed at room temperature. In the case of 800 MeV Xe and 320 MeV Au irradiations, bimodal splayed CDs were introduced by controlling the angle between the beam and the crystal \emph c-axis.
Magnetization measurements were performed by using a commercial SQUID magnetometer (MPMS-XL5, Quantum Design). \emph J${\rm {_c}}$ [A/cm${\rm {^2}}$] is estimated by using the extended Bean model \cite{bean1964magnetization}, which is
\begin{equation}
\hspace{3.5cm} J_c = \frac{{\mathit {\rm {20}}{\Delta} M}}{{a\left(1 - \frac{a}{{3b}}\right)}} \hspace{0.5cm} ({a<b}),
\end{equation}
where ${\mathit{\Delta} M}$ [emu/cm${\rm {^3}}$] is the difference of magnetization when the applied field is swept down and up. \emph a [cm] and \emph b [cm] are width and length of the sample, respectively. We also made detailed angular dependent measurements by using a miniature Hall probe (HG-0711, AsahiKASEI, 1.2×0.5×0.3 mm${\rm {^3}}$) with an  active area of 40$\times$40 $\mu$m${^2}$. For such Hall probe measurements, we modified the horizontal sample rotator for MPMS so that two Hall probes can be placed on the rotating stage, one for the magnetic induction measurements of samples and another for the measurements of external field. Schematic diagram of the top view and side view of the rotating stage is shown in Figs. \ref{fig:1}(a) and (b), respectively. Using this rotator with miniature Hall probes, the direction of the magnetic field can be changed continuously without taking the sample out of the MPMS system. The Hall resistance was measured by an AC resistance bridge (LR700). The local induction measured by the Hall probe is the combination of the applied magnetic field and the field generated by the current in the sample. The field generated from the sample is usually much smaller compared with the external field. Furthermore, the Hall resistance of Hall probe as a function of external field is not strictly linear at low temperatures and high magnetic fields. To obtain the reliable value of the local induction generated by the sample, it is necessary to obtain a reliable calibration curve. In our case, we first measured the Hall resistance with a magnetic field sequence of 0 Oe→20 kOe→-20 kOe→0 Oe at \emph T = 10 K, where the effect of superconductivity is absent. For the evaluation of the local induction measured at a given field, we interpolated the calibration curve after proper rescaling of the signal to take into account the temperature dependence of Hall resistance. This \emph{in situ} experiment allows us efficient measurements on the sample with many different directions of the applied magnetic field.

\section{\label{sec:level1}Results and Discussion}
\subsection{\label{sec:level2}O-DO transition-induced peak effect}
\begin{figure}[b]
\flushleft
\includegraphics[scale=0.57]{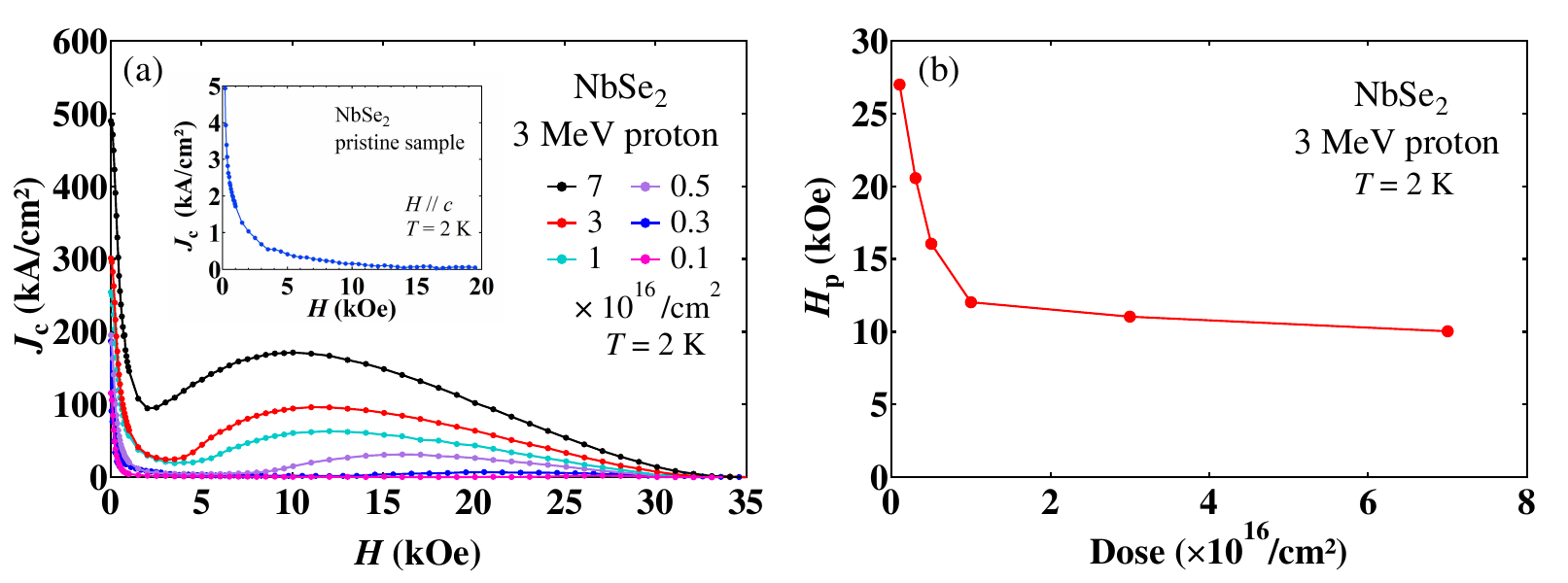}
\caption{\label{fig:2}(a) Magnetic field dependence of \emph J${\rm {_c}}$ for a pristine and 3 MeV proton-irradiated NbSe${\rm {_2}}$ single crystals with different doses at \emph T = 2 K. (b) Dose dependence of the peak field for NbSe${\rm {_2}}$ irradiated by 3 MeV protons at \emph T = 2 K.}
\end{figure}

Figure \ref{fig:2}(a) shows magnetic field dependence of \emph J${\rm {_c}}$ for a pristine and 3 MeV proton-irradiated NbSe${\rm {_2}}$ single crystals with different doses. Other than the \emph J${\rm {_c}}$ enhancement with increasing irradiation dose, clear peak effects were observed. In type II superconductors, peak effects located at high fields close to \emph H${\rm {_c{_2}}}$ is frequently observed, which is originated from the transition between the ordered state and the disordered state of vortices at an applied field of \emph H${\rm {_p}}$ \cite{paltiel2000instabilities}. When the stability of these phases is affected by perturbation to the system, the peak also changes accordingly. For example, introducing point defects into a superconductor stabilize the disordered phase, making the transition from the ordered state to disordered state occur at lower fields. Namely, \emph H${\rm {_p}}$ shifts from high fields to low fields as disorder increases. Point defects introduced by 3 MeV proton irradiation are known to enhance disordered phase \cite{moore2005angular}. As shown in Fig. \ref{fig:2}(b), \emph H${\rm {_p}}$ monotonically decreases with increasing proton irradiation dose. This is consistent with the understanding of the feature for O-DO transition-induced peak effect.

\begin{figure}[!h]
\centering
\includegraphics[scale=0.52]{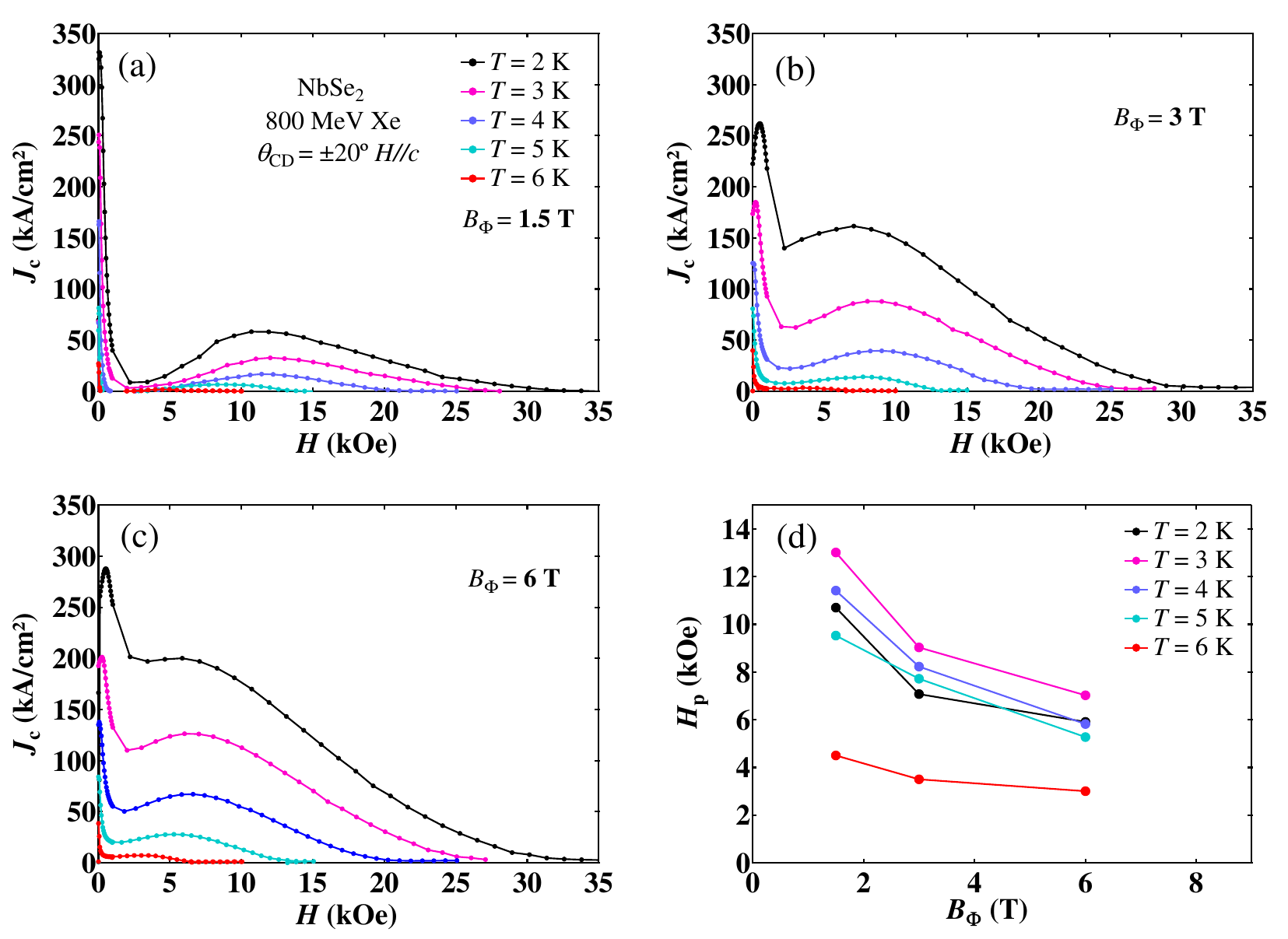}
\caption{\label{fig:3}Magnetic field dependence of \emph J${\rm {_c}}$ for NbSe${\rm {_2}}$ with splayed CDs at $\theta_{\mathrm{CD}}$ = ±20° with total \emph B{${_\Phi}$} of (a) 1.5 T, (b) 3 T, and (c) 6 T introduced by 800 MeV Xe irradiation. (d) \emph B{${_\Phi}$} dependence of \emph H${\rm {_p}}$ induced by O-DO transition for NbSe${\rm {_2}}$ with splayed CDs ($\theta_{\mathrm{CD}}$ = ±20°) introduced by 800 MeV Xe irradiation at different temperatures.}
\end{figure}
\begin{figure*}
\centering
\includegraphics[scale=0.65]{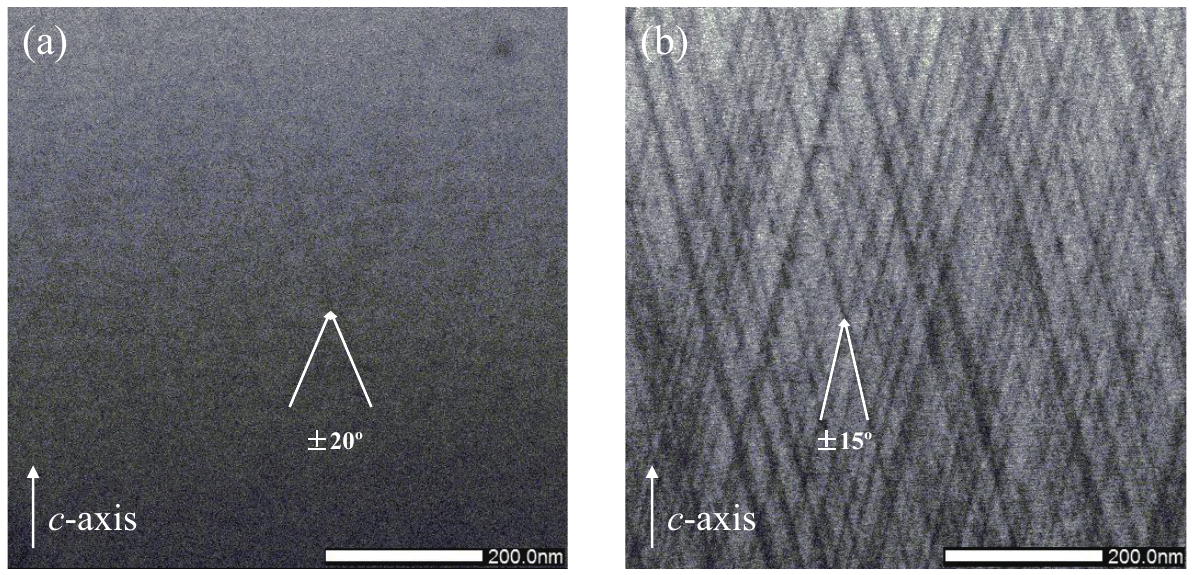}
\caption{\label{fig:4}TEM images of NbSe${\rm {_2}}$ irradiated by (a) 800 MeV Xe ($\theta_{\mathrm{CD}}$ = ±20°, \emph B{${_\Phi}$} = 1.5 T+ 1.5 T) and (b) 320 MeV Au ($\theta_{\mathrm{CD}}$ = ±15°, \emph B{${_\Phi}$} = 2 T+ 2 T). Continuous CDs are introduced by 320 MeV Au irradiation, while introduced are faint in 800 MeV Xe irradiated sample.}
\end{figure*}
\begin{figure*}
\flushleft
\includegraphics[scale=0.55]{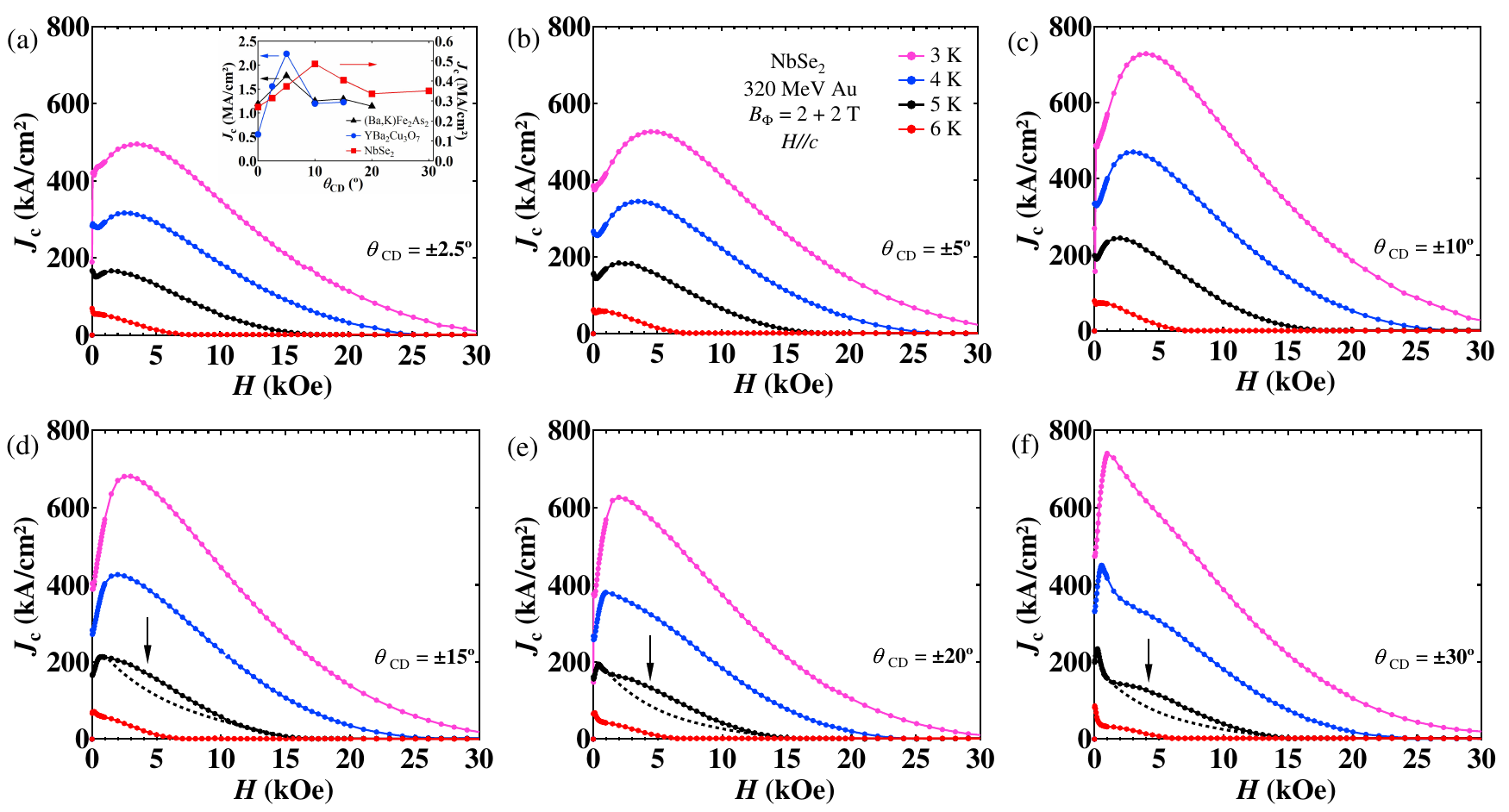}
\caption{\label{fig:5}Magnetic field dependence of \emph J${\rm {_c}}$ for NbSe${\rm {_2}}$ with splayed CDs at (a) $\theta_{\mathrm{CD}}$ = ±2.5°, (b) $\theta_{\mathrm{CD}}$ = ±5°, (c) $\theta_{\mathrm{CD}}$ = ±10°, (d) $\theta_{\mathrm{CD}}$ = ±15°, (e) $\theta_{\mathrm{CD}}$ = ±20°, and (f) $\theta_{\mathrm{CD}}$ = ±30° with \emph B{${_\Phi}$} = 2 T + 2 T introduced by 320 MeV Au irradiation. The dashed line in (d), (e), and (f) shows a possible magnetic field dependence of \emph J${\rm {_c}}$ without point-pinning-like contributions of CDs. Black arrows in (d), (e), and (f) indicate the hump related to O-DO transition-induced peak effect. The inset of (a) shows \emph J${\rm {_c}}$ at \emph H = 10 kOe as a function of ${|}$$\theta_{\mathrm{CD}}$${|}$ for the samples with splayed CDs shown in (a)-(f). As a reference, \emph J${\rm {_c}}$ at \emph H = 10 kOe for YBa${_2}$Cu${_3}$O${_7}$ \cite{krusin1996enhanced}\ and (Ba,K)Fe${_2}$As${_2}$ \cite{park2018field}\ with splayed CDs as a function of ${|}$$\theta_{\mathrm{CD}}$${|}$ is also plotted.}
\end{figure*}
\begin{figure}
\centering
\includegraphics[scale=0.8]{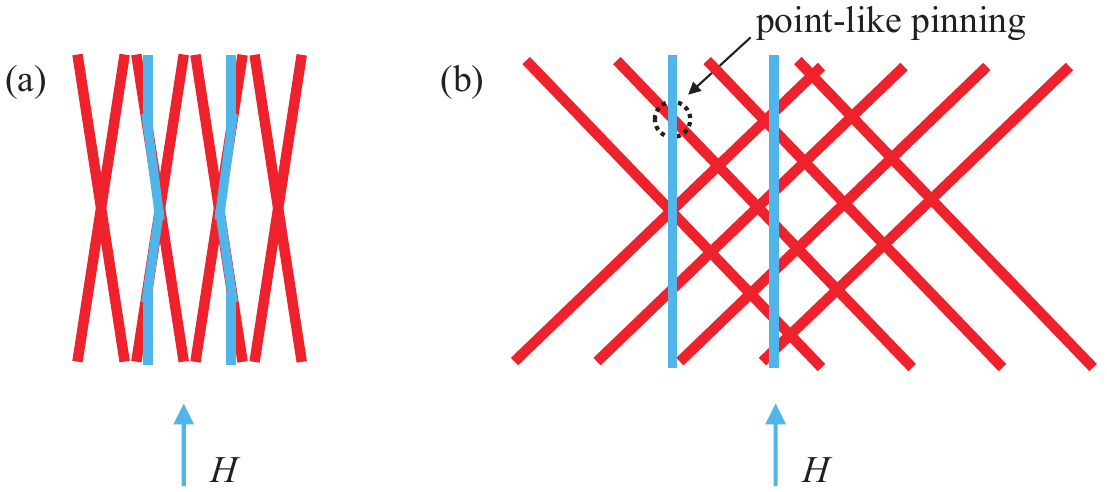}
\caption{\label{fig:6}Two different types of pinning of vortices (blue lines) by splayed CDs (red lines) with different $\left| \theta_{\mathrm{CD}} \right|$. (a) At small $\left| \theta_{\mathrm{CD}} \right|$, larger portions of vortices are trapped on CDs, making the pinning mostly columnar. (b) At larger $\left| \theta_{\mathrm{CD}} \right|$, only short segments of vortices are trapped by CDs, making the pinning point-like.}
\end{figure}
\begin{figure}
\centering
\includegraphics[scale=0.8]{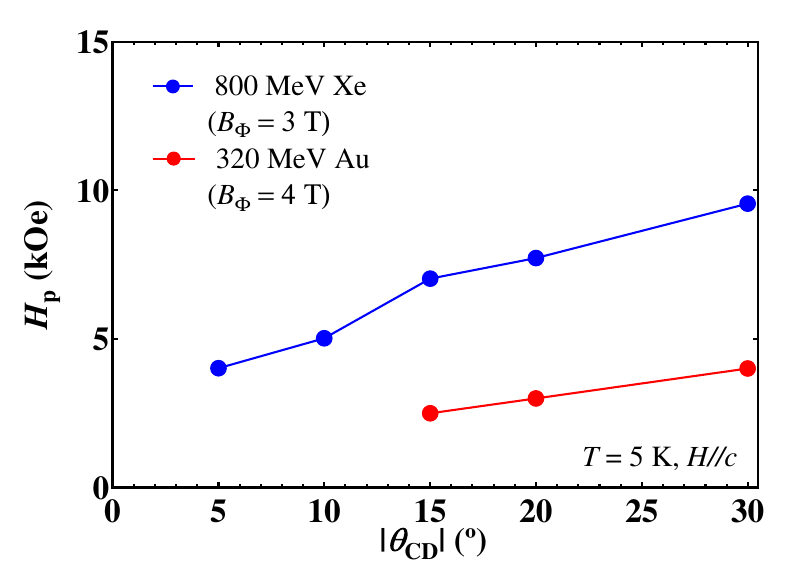}
\caption{\label{fig:7}$\left| \theta_{\mathrm{CD}} \right|$ dependence of \emph H${\rm {_p}}$ observed at \emph T = 5 K with \emph H//\emph c-axis for NbSe${\rm {_2}}$ single crystals with splayed CDs introduced by 320 MeV Au irradiation (Total \emph B{${_\Phi}$} = 4 T) and Xe irradiation (Total \emph B{${_\Phi}$} = 3 T).}
\end{figure}

Figures \ref{fig:3}(a)-(c) show the \emph J${\rm {_c}}$ as a function of magnetic field at various temperatures in NbSe${\rm {_2}}$ with splayed CDs introduced by 800 MeV Xe irradiation. In this data set, $\left| \theta_{\mathrm{CD}} \right|$ was held constant at 20° and only the total \emph B{${_\Phi}$} was changed from 1.5 T to 6 T. It can be seen that the peak located in the middle field region shifts from high fields to low fields with increasing \emph B{${_\Phi}$} at all temperatures as shown in the \emph H${\rm {_p}}$-\emph B{${_\Phi}$} plot in Fig. \ref{fig:3}(d). This trend is similar to what we observed in the proton-irradiated NbSe${\rm {_2}}$, where the peak shifts to lower fields with increasing proton irradiation dose. Such an observation indicates a possible formation of the O-DO transition-induced peak effect in superconductors with splayed CDs. The formation of CDs by heavy ions can be confirmed by transmission electron microscopy (TEM) results as shown in Figs. \ref{fig:4}(a) and (b). The difference between the splayed CDs introduced by 800 MeV Xe and 320 MeV Au ions in NbSe${\rm {_2}}$ single crystals is that in the former case, only some faint traces of CDs are introduced while in the latter case distinct continuous CDs are created. In the case of 320 MeV Au irradiation, the introduced CDs in NbSe${\rm {_2}}$ single crystals have diameters $\sim$6-8 nm. This value is comparable to the in-plane coherence length of NbSe${\rm {_2}}$ of \emph {$\xi$}${_a{_b}}$${\approx}$ 7.4 nm, and introduced CDs are expected to be effective pinning centers for vortices.

Figures \ref{fig:5}(a)-(f) show the magnetic field dependence of \emph J${\rm {_c}}$ for NbSe${\rm {_2}}$ with splayed CDs introduced by 320 MeV Au irradiation. Here, the total \emph B{${_\Phi}$} was held constant at 4 T, and evolution of the peak effect with $\left| \theta_{\mathrm{CD}} \right|$ was studied. Pronounced peak effects can be observed for all samples. With increasing splayed angles, the broad peak at intermediate fields gradually shifts from low fields to high fields. The hump-like feature at intermediate fields in samples with $\left| \theta_{\mathrm{CD}} \right|$ ${\geq}$ 15° may be induced by O-DO transition of vortices similar to the case of 800 MeV Xe irradiation, which can be explained by the formation of point-like pinning induced by partial pins. In NbSe${\rm {_2}}$ with splayed CDs with small $\left| \theta_{\mathrm{CD}} \right|$, most parts of vortices can be trapped by CDs as shown in Fig. \ref{fig:6}(a). In this case, CDs act as correlated defects. As $\left| \theta_{\mathrm{CD}} \right|$ increases, portions of vortices trapped by CDs become small as shown in Fig. \ref{fig:6}(b). The effective length of vortices trapped by CDs is ${\sim\frac{2\xi}{\sin(\theta_{\mathrm{CD}})}}$, where $\xi$ is the coherence length of the superconductor. It means that the trapped volume of vortices by CDs decreases with increasing $\left| \theta_{\mathrm{CD}} \right|$, and the pinning becomes more like point pinning. Figure \ref{fig:7} shows the \emph H${\rm {_p}}$ as a function of $\left| \theta_{\mathrm{CD}} \right|$ for NbSe${\rm {_2}}$ with splayed CDs generated by 800 MeV Xe and 320 MeV Au irradiations at \emph T = 5 K. It is clear that \emph H${\rm {_p}}$ shifts from low fields to high fields with increasing $\theta_{\mathrm{CD}}$ in both cases. Here, we make a rough estimation of the density of point-like pinning centers in the case of splayed CDs as follows. We consider a sample with a thickness of 10 ${\mu}$m with tilted (splayed) CD with $|\theta_{\mathrm{CD}}$${|}$ = 10°. At an applied external field of \emph H = 10 kOe along the \emph c-axis, the average spacing of vortices is ${\sim}$50 nm. In this case, one tilted (splayed) CD can cross ${\sim}$35 vortices at the maximum. When the density of tilted (splayed) vortices is \emph B{${_\Phi}$} = 4 T, the maximum areal density of crossing points of tilted (splayed) CDs and vortices is calculated as 35 $\times$ 2 $\times$ 10${^1{^1}}$/cm${^2}$= 7 $\times$ 10${^1{^2}}$/cm${^2}$. It should be noted that the actual density of point-pinning centers due to tilted (splayed) CDs can be smaller than this. Even this maximum density of point-like pinning centers is 3 orders of magnitude smaller than the typical value of proton dose of 1 $\times$ 10${^1{^6}}$/cm${^2}$. The fact that point-like pinning due to CDs with large $\theta_{\mathrm{CD}}$ can cause similar O-DO transition due to proton irradiation at similar fields can be reconciled by the fact that O-DO transition is determined by both the density and strength of point-like pinning centers. In the case of tilted (splayed) CDs crossing with a vortex parallel to the \emph c-axis, the effective dimensions of a point-like pinning center are given by ${\sim}$\emph d $\times$ \emph d $\times$ 2${\xi}$/tan($\theta_{\mathrm{CD}}$), where \emph d is the diameter of CDs and ${\xi}$ is the coherence length. Both \emph d and ${\xi}$ are 6-8 nm for NbSe${\rm {_2}}$, the effective dimensions of a point-like pinning center are much larger than those for point defects created by proton irradiation, although the exact dimension of these point defects are not well known. The qualitative difference in the shapes of \emph J${\rm {_c}}$-\emph H curves for 320 MeV Au, 800 MeV Xe, and 3 MeV proton irradiations can be explained by the background pinning induced by CDs. Namely, background pinning is the strongest in samples irradiated by 320 MeV Au, weak in 800 MeV Xe irradiation, and absent in 3 MeV proton irradiation. The \emph J${\rm {_c}}$ close to the peak field (peaks are indicated by black arrow) may be the superposition of effects from point defects and CDs (dashed lines in Figs. \ref{fig:5}(d)-(f)). 

\subsection{\label{sec:level2}Self-field peak effect in NbSe${\rm {_2}}$ with splayed CDs}
Magnetic field dependence of \emph J${\rm {_c}}$ for NbSe${\rm {_2}}$ with splayed CDs introduced by 800 MeV Xe irradiation are shown in Figs. \ref{fig:3}(a)-(c), which indicate the presence of two kinds of peaks under certain conditions. Two peaks can be clearly identified in the sample with \emph B{${_\Phi}$} = 3 T, $\theta_{\mathrm{CD}}$ = ±20° at \emph T = 2 K. Broad peaks (hump) located at intermediate fields are caused by O-DO transition of vortices as discussed in section 3.1. In addition, another sharp peaks are observed close to the zero field. These peaks are originated from the self-field effect, which will be further discussed together with similar peaks in NbSe${\rm {_2}}$ with splayed CDs introduced by 320 MeV Au irradiation.
\begin{figure}[t]
\centering
\includegraphics[scale=0.75]{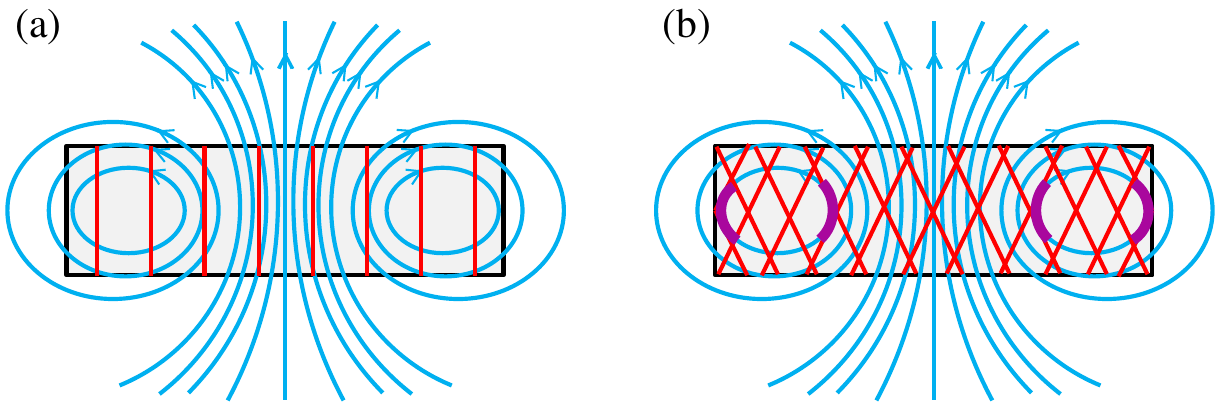}
\caption{\label{fig:8}Schematic figures illustrating the distribution of magnetic lines of force near the superconductor in the remanent state with (a) parallel CDs and (b) splayed CDs. Red and blue lines represent CDs and magnetic lines of force. The purple regions in (b) indicate the parts of vortices pinned by CDs.}
\end{figure}

In Figs. \ref{fig:5}(d)-(f), magnetic field dependence of \emph J${\rm {_c}}$ for NbSe${\rm {_2}}$ with splayed CDs introduced by 320 MeV Au irradiation also indicate the presence of two kinds of peaks under certain conditions. Sharp peaks are clearly observed at low fields in the sample with \emph B{${_\Phi}$} = 4 T and $\theta_{\mathrm{CD}}$ = ±30° below \emph T = 5 K. At 5 K, the sharp peak near zero field for this sample shows up at \emph H${\rm {_p}}$ $\sim\rm{0.2}$ kOe. This \emph H${\rm {_p}}$ value is very close to the self-field \emph H${\rm {_s{_f}}}$ = 0.2 kOe for this sample. Therefore, we can consider that this low-field peak has similar mechanism to the self-field peak effect which is frequently observed in superconductors with parallel CDs \cite{tamegai2012effects}. The self-field peak effect in thin superconductors with parallel CDs is induced by the curvature of magnetic flux lines. When the external magnetic field is lower than \emph H${\rm {_s{_f}}}$ in thin superconductors, the magnetic flux lines near the edge of the sample are curved significantly, which results in ineffective pinning by parallel CDs and causes a reduction of \emph J${\rm {_c}}$ below \emph H${\rm {_s{_f}}}$. When the external magnetic field is increased, the curved flux lines are straightened making vortices being effectively pinned by CDs \cite{tamegai2012effects}. Hence, \emph J${\rm {_c}}$ reaches a maximum at $\sim $\emph H${\rm {_s{_f}}}$. In this scenario, \emph J${\rm {_c}}$ at the center of the superconductor under zero field is larger than that at the edge. Since vortices at the center of a superconductor are aligned with the direction of the external magnetic field, they can be effectively pinned by parallel CDs as schematically shown in Fig. \ref{fig:8}(a). Such pinning can result in a larger \emph J${\rm {_c}}$ in the central region than that near the edge of the superconductor. In superconductors with splayed CDs with large $\left| \theta_{\mathrm{CD}} \right|$, the formation mechanism of the self-field peak effect can be different. In such superconductors, a part of curved magnetic flux lines near the sample edge can be effectively pinned by splayed CDs under the self-field as shown by the purple region in Fig. \ref{fig:8}(b). At a magnetic field slightly less than \emph H = \emph H${\rm {_s{_f}}}$, \emph J${\rm {_c}}$ near the edge of the superconductor can be larger than that in the center of the superconductor. Such a high \emph J${\rm {_c}}$ region forms a closed loop close to the edge of the sample. As the external field is reduced towards zero, the spanned area by the closed loop with larger \emph J${\rm {_c}}$ shrinks, leading to the suppression of the measured magnetization. Salient feature of the peak effect at low fields in samples with splayed CDs is the sharpness of the peak compared to that in samples with parallel CDs \cite{tamegai2012effects}. Qualitatively, compared with the formation of the self-field peak effect observed in samples with parallel CDs, the curved magnetic flux lines at the edge of superconductors with splayed CDs could be effectively pinned before they are fully straightened. In such a case, the self-field effect in samples with splayed CDs may occur at lower magnetic fields, which can be one of the reasons to induce a sharper self-field peak feature.
\begin{figure}[b]
\centering
\includegraphics[scale=0.8]{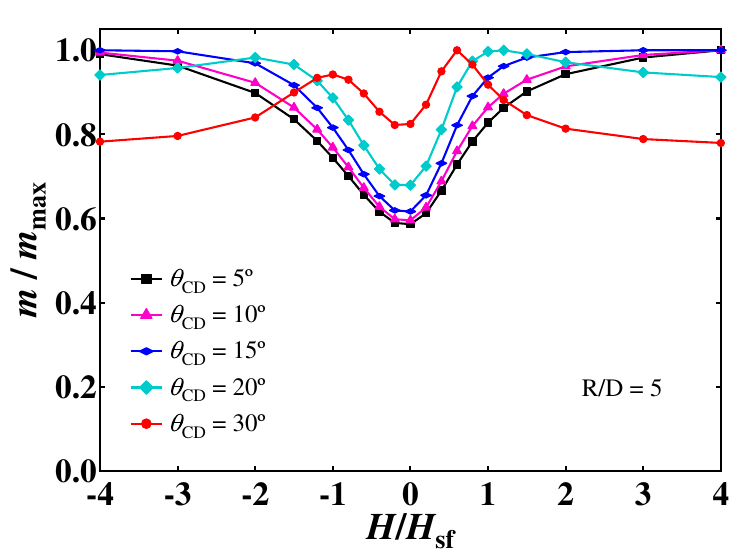}
\caption{\label{fig:9}Simulation results of the normalized magnetic moment (normalized by the maximum moment) as a function of the normalized magnetic field (normalized by the self-field) for disk-shaped superconductors with different $\left| \theta_{\mathrm{CD}} \right|$.}
\end{figure}

To further understand the mechanism of the self-field peak effect in superconductors with splayed CDs, numerical calculations are performed. We calculated the field dependence of magnetic moment of a disk-shaped superconductor where the \emph J${\rm {_c}}$ is considered to be a function of the angle between local magnetic induction and the \emph c-axis. The magnetic field generated by a circular current at a certain point can be obtained by the integration of Eqs. (2) and (3) over the whole volume as described in Ref. \cite{smythe1950static}, where \emph H${ {_r}}$ and \emph H${ {_z}}$ are the radial and axial components of the magnetic field generated in cylindrical coordinates, \emph r and \emph z are arbitrary coordinates, \emph a is the radius of the current \emph I, and \emph K(\emph k) and \emph E(\emph k) are the first and second kinds of complete elliptic integrals, respectively, with \emph k given by Eq. (4). The magnetization can be calculated by the integration of magnetic moment of current loops over the whole volume. The extended midpoint rule was selected to perform the numerical integrations of these two equations similar to the treatment in Ref. \cite{conner1991calculations}. The \emph J${\rm {_c}}$(\emph {${\theta}$}), which is determined by the angle \emph {${\theta}$} of the local magnetic induction from the \emph c-axis, is set as a function of the angle \emph {${\theta}$}${\rm {_C{_D}}}$ in the form consisting of two Gaussian functions centered at +\emph {${\theta}$}${\rm {_C{_D}}}$ and -\emph {${\theta}$}${\rm {_C{_D}}}$ as described by Eq. (5). In the calculation, we set \emph J${\rm {_c{_0}}}$ = \emph J${\rm {_c{_1}}}$ and ${\mathit{\Delta\theta}}$ = 20° for all \emph {${\theta}$}${\rm {_C{_D}}}$, and ratio of the radius (\emph R) and thickness (\emph D) of the disk as \emph R/\emph D = 5.
\begin{equation}
H_r(a,r,z) = \frac{I}{2\pi r}\cdot\frac{z}{\sqrt{(a+r)^2+z^2}}\cdot\left[-K(k) + \frac{a^2+r^2+z^2}{(a-r)^2+z^2}\cdot E(k)\right]
\end{equation}
\begin{equation}
H_z(a,r,z) = \frac{I}{2\pi}\cdot\frac{1}{\sqrt{(a+r)^2+z^2}}\cdot\left[K(k) + \frac{a^2-r^2-z^2}{(a-r)^2+z^2}\cdot E(k)\right]
\end{equation}
\begin{equation}
\hspace{3cm} k = \left[\frac{4ar}{(a+r)^2+z^2}\right]^{\frac{1}{2}}
\end{equation}
\begin{equation}
\hspace{0cm} J_{\mathrm{c}}(\emph {${\theta}$}) = J_{\mathrm{c0}} + J_{\mathrm{c1}} \left(\exp\left(-\left(\frac{\theta - \theta_{\mathrm{CD}}}{\mathit{\Delta\theta}}\right)^2\right) + \exp\left(-\left(\frac{\theta + \theta_{\mathrm{CD}}}{\mathit{\Delta\theta}}\right)^2\right)\right)
\end{equation}
\begin{figure*}[b]
\flushleft
\includegraphics[scale=0.52]{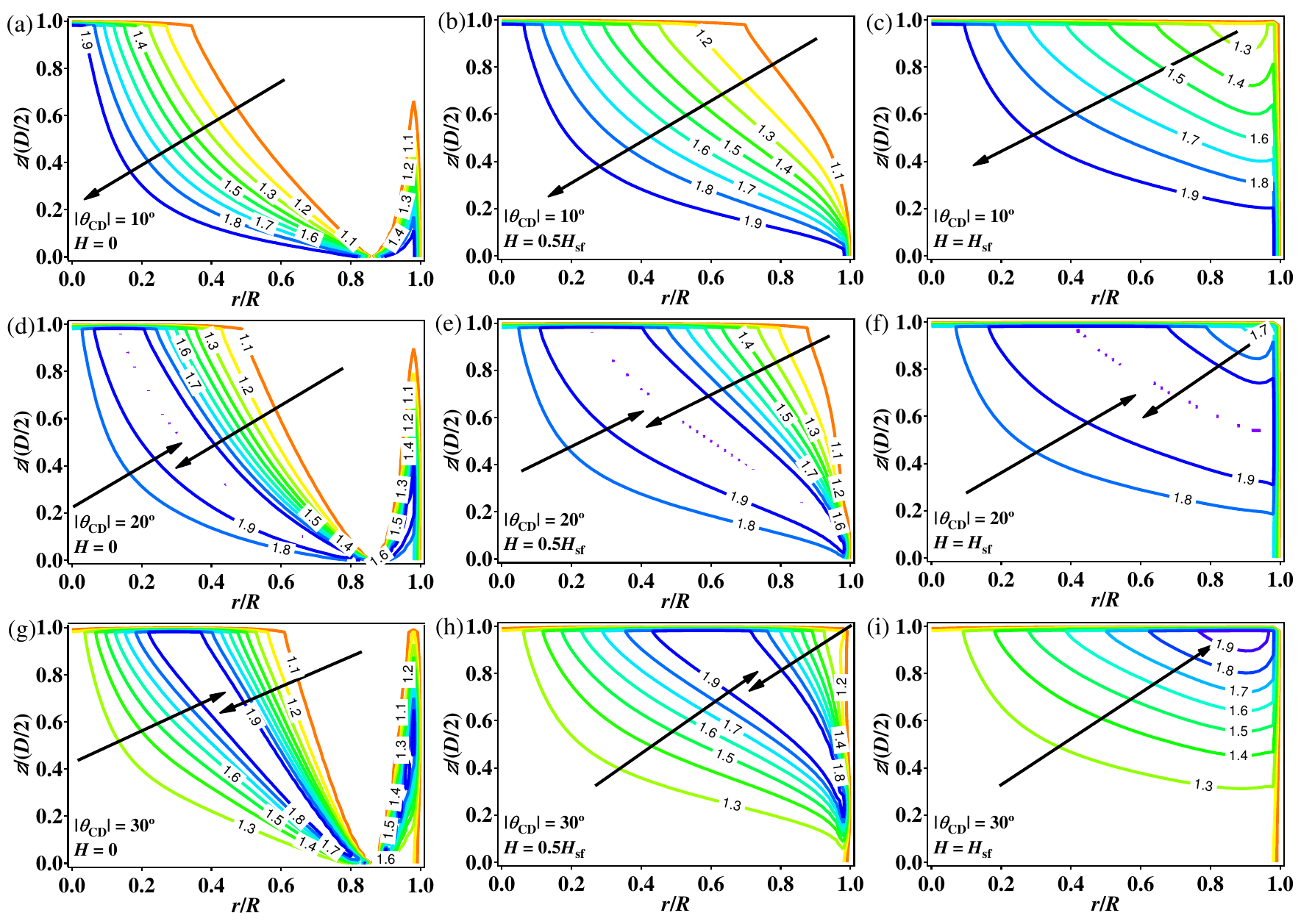}
\caption{\label{fig:10}\emph J${\rm {_c}}$ distributions calculated for disk-shaped superconductors with angle dependent \emph J${\rm {_c}}$ given by Eq. (5) with $\left| \theta_{\mathrm{CD}} \right|$ = 10° ((a)-(c)), 20° ((d)-(f)), and 30° ((g)-(i)) at \emph H = 0 ((a), (d), (g)), \emph H = \emph H${\rm {_s{_f}}}$ ((b), (e), (h), and \emph H = 2\emph H${\rm {_s{_f}}}$ ((c), (f), (i)). (Arrows indicate the direction of \emph J${\rm {_c}}$ increase.)}
\end{figure*}

Figure \ref{fig:9} illustrates the normalized magnetic moment (normalized by the maximum moment) as a function of normalized magnetic field (normalized by the self-field) for disk-shaped superconductors with effects from various angles of splayed CDs. Notably, it can be observed that a sharp peak appears at large $\left| \theta_{\mathrm{CD}} \right|$, which is consistent with the experimental observations in Fig. \ref{fig:5}. It should be noted that peaks do not show up at smaller $\left| \theta_{\mathrm{CD}} \right|$, since the magnetic field dependence of \emph J${\rm {_c}}$ is not taken into account. The \emph J${\rm {_c}}$ distributions in samples with different $\left| \theta_{\mathrm{CD}} \right|$ under the application of external fields of 0, \emph H${\rm {_s{_f}}}$, and 2\emph H${\rm {_s{_f}}}$ are depicted in Figs. \ref{fig:10}. Two distinct regions with high \emph J${\rm {_c}}$ are observed at \emph H = 0 for all three samples. For the sample with $\left| \theta_{\mathrm{CD}} \right|$ = 10°, one high \emph J${\rm {_c}}$ region is located at the center, while another high \emph J${\rm {_c}}$ region is found near the edge of the sample. In samples with $\left| \theta_{\mathrm{CD}} \right|$ = 20° and 30°, the high \emph J${\rm {_c}}$ region near the edge are observed and becomes even prominent, and another high \emph J${\rm {_c}}$ region is observed in the middle region. These observations indicate the formation of non-uniform \emph J${\rm {_c}}$ flow, as schematically depicted in Fig. \ref{fig:8}.

For proper understanding of the shift of the peak field, it is necessary to take into account the field dependence of \emph J${\rm {_c}}$. As shown in Figs. \ref{fig:10}, for the sample with $\left| \theta_{\mathrm{CD}} \right|$ = 10°, increasing the applied field leads to an expansion of the large \emph J${\rm {_c}}$ region at the center, indicating a monotonic increase in magnetic moment. By contrast, in samples with $\left| \theta_{\mathrm{CD}} \right|$ = 20° and 30°, as the external field increases, the high \emph J${\rm {_c}}$ regions are expanded, signifying an increase in magnetic moment. However, at the same time, the low \emph J${\rm {_c}}$ region at the center expands. Considering these factors, it is plausible to conclude that the magnetic moment for these two samples form a maximum value with increasing external field.

 \subsection{\label{sec:level2}Anomalous peak effect in NbSe$_2$ with splayed CDs}
\begin{figure}[htb]
\centering
\includegraphics[scale=0.6]{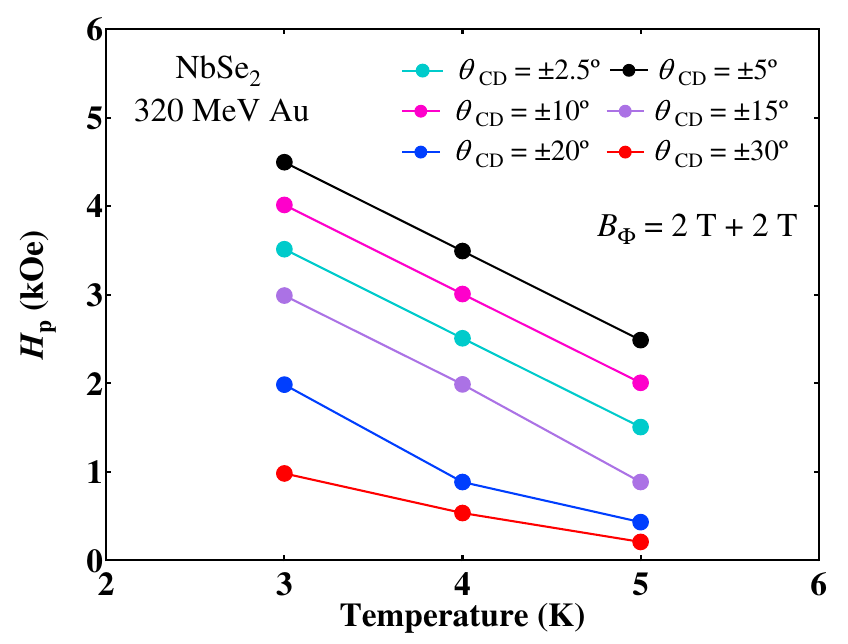}
\caption{\label{fig:11}Temperature dependence of the peak field \emph H${\rm {_p}}$ for NbSe${\rm {_2}}$ single crystals with splayed CDs introduced by 320 MeV Au irradiation at various $\left| \theta_{\mathrm{CD}} \right|$ with \emph B{${_\Phi}$} = 2 T + 2 T.}
\end{figure}

In Figs. \ref{fig:5}(a)-(f), other than the self-field peak effects located at low fields as described in section 3.2, an anomalous peak effect was observed at intermediate fields. As mentioned in the introduction, anomalous peak effects are frequently observed at \emph H = \emph H${\rm {_p}}$ $\sim$1/3\emph B{${_\Phi}$} in superconductors with splayed CDs. However, in the present study on NbSe${\rm {_2}}$ with splayed CDs, the value of \emph H${\rm {_p}}$/\emph B{${_\Phi}$} is not constant. Figure \ref{fig:11} is the temperature dependence of \emph H${\rm {_p}}$ for samples with different $\left| \theta_{\mathrm{CD}} \right|$. \emph H${\rm {_p}}$ decreases linearly as a function of temperature for all six samples. It is also clear that the value of \emph H${\rm {_p}}$ varies with $\theta_{\mathrm{CD}}$. In a previous study on the effect of \emph B{${_\Phi}$} on the anomalous peak effect \cite{tamegai2023anomalous}, it was found that \emph H${\rm {_p}}$ initially increases with increasing \emph B{${_\Phi}$} and finally decreases as \emph B{${_\Phi}$} becomes too large until it was completely suppressed at large enough \emph B{${_\Phi}$}. For samples with different \emph B{${_\Phi}$} and $\theta_{\mathrm{CD}}$ under several temperatures, the value of \emph H${\rm {_p}}$/\emph B{${_\Phi}$} varies in the range of 0.1 to 0.7 \cite{tamegai2023anomalous}. In superconductors with random distribution of CDs, the pinning force density (\emph F${\rm {_p}}$) reaches a maximum value when all vortices are pinned by CDs. In other words, \emph F${\rm {_p}}$ is expected to reach the maximum value when the applied magnetic field is close to \emph B{${_\Phi}$}, above which it stays constant. So, anomaly in \emph J${\rm {_c}}$-\emph H is expected at \emph H/\emph B{${_\Phi}$} $\sim$ 1. However, several factors can affect the field value for the maximum \emph J${\rm {_c}}$, such as interactions between vortices \cite{blatter1994vortices}, formation of vortex entanglement \cite{hwa1993flux}, and the influence of interlayer phase coherence in layered superconductors \cite{shibauchi1999interlayer}, which cause the anomaly to occur at lower \emph H/\emph B{${_\Phi}$}. As mentioned above, it was reported that \emph H${\rm {_p}}$/\emph B{${_\Phi}$} changes considerably as a function of total \emph B{${_\Phi}$} in NbSe${\rm {_2}}$ with splayed CDs \cite{tamegai2023anomalous}. Even in the case of smallest total \emph B{${_\Phi}$} of 0.2 T that have ever been reported, the peak appears at \emph H${\rm {_p}}$ = 1.4 kOe, making \emph H${\rm {_p}}$/\emph B{${_\Phi}$} = 0.7 \cite{tamegai2023anomalous}. However, it should be noted that the anomalous peak effect is not observed in the limit of \emph B{${_\Phi}$} = 0 T, namely, in the pristine sample. Furthermore, the $\theta_{\mathrm{CD}}$ value for the maximum \emph J${\rm {_c}}$ observed in NbSe${\rm {_2}}$ is different compared to other unconventional superconductors. In YBa${_2}$Cu${_3}$O${_7}$ \cite{krusin1996enhanced} and (Ba,K)Fe${_2}$As${_2}$ \cite{park2018field} the optimal angle for the maximum \emph J${\rm {_c}}$ is reported to be 5°, while it is 10° in NbSe${\rm {_2}}$ as shown in the inset of Fig. \ref{fig:5}(a).
\begin{figure}[htb]
\centering
\includegraphics[scale=0.6]{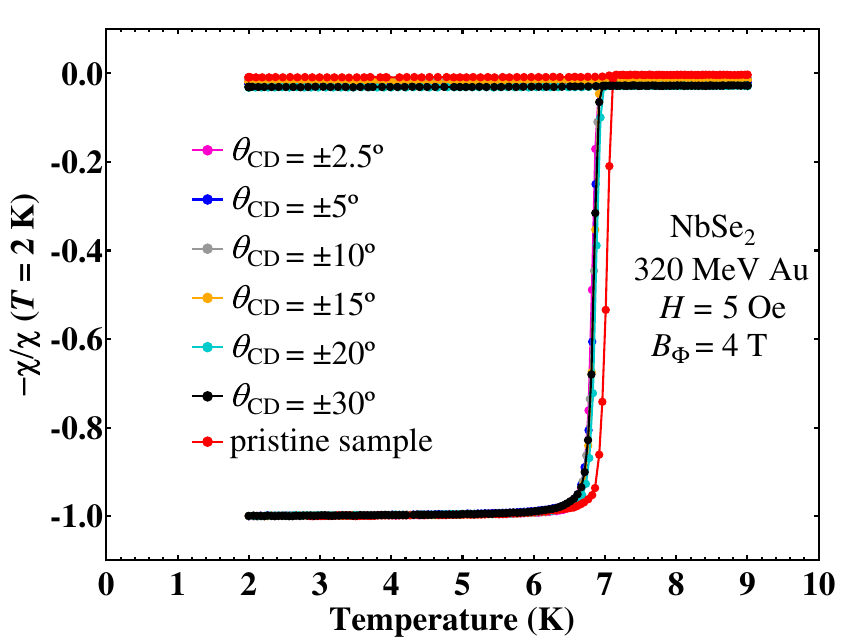}
\caption{\label{fig:12}Temperature dependence of the normalized magnetic susceptibility at \emph H = 5 Oe for a pristine NbSe${\rm {_2}}$ single crystal and those with splayed CDs introduced by 320 MeV Au irradiation at various $\left| \theta_{\mathrm{CD}} \right|$ with \emph B{${_\Phi}$} = 2 T + 2 T.}
\end{figure}

As reported in the previous study on NbSe${\rm {_2}}$ with parallel CDs, \emph T${\rm {_c}}$ decreases monotonically with increasing \emph B{${_\Phi}$} \cite{li2022suppression}. Actually, \emph T${\rm {_c}}$ of NbSe${\rm {_2}}$ is influenced by various factors such as charge-density wave, disorder, and lattice expansion \cite{li2022suppression,cho2018using,moulding2020absence,majumdar2020interplay}. To check whether different \emph T${\rm {_c}}$ caused by different $\left| \theta_{\mathrm{CD}} \right|$ can affect the peak effects, we have measured the \emph M-\emph T curves for all 320 MeV Au irradiated samples with \emph B{${_\Phi}$} = 2 T + 2 T and different $\left| \theta_{\mathrm{CD}} \right|$ as shown in Fig. \ref{fig:12}. As a reference, \emph M-\emph T curve for a pristine sample is also included in Fig. \ref{fig:12}. Obviously, all \emph M-\emph T curves overlap to each other, indicating that their \emph T${\rm {_c}}$ does not change in samples with different $\left| \theta_{\mathrm{CD}} \right|$ as long as the total \emph B{${_\Phi}$} is the same. In the \emph J${\rm {_c}}$-\emph H curves for NbSe${\rm {_2}}$ with splayed CDs introduced by 320 MeV Au irradiation as shown in Fig. \ref{fig:5}, we observed three kinds of peak effects. For samples with $\theta_{\mathrm{CD}}$ = ±2.5°, ±5°, and ±10°, the observed peaks at all measured temperatures are very similar. The shape of these peaks is broad and similar to the anomalous peak effect observed in Ba${\rm {_1{_-}}}$$_x$K$_x$Fe$_2$As$_2$ \cite{park2018field}. In addition, \emph H${\rm {_p}}$ values of these peaks are larger than the self-field (\emph H${\rm {_s{_f}}}$(\emph T = 2 K) for all these samples are less than 1 kOe). For samples with $\left| \theta_{\mathrm{CD}} \right|$ \textgreater{} 10°, the peak becomes sharper as $\left| \theta_{\mathrm{CD}} \right|$ increases. In samples with $\theta_{\mathrm{CD}}$ = ±15° and ±20°, the shapes of the peaks observed at 3 K and 4 K are still similar to that of the anomalous peak effect. However, as the temperature increases, a broad peak splits into two features as seen in \emph J${\rm {_c}}$-\emph H curve at \emph T = 5 K. Coexistence of two features is most clear in the sample with $\theta_{\mathrm{CD}}$ = ±30° at \emph T = 3-5 K.

\begin{figure}
\centering
\includegraphics[scale=0.7]{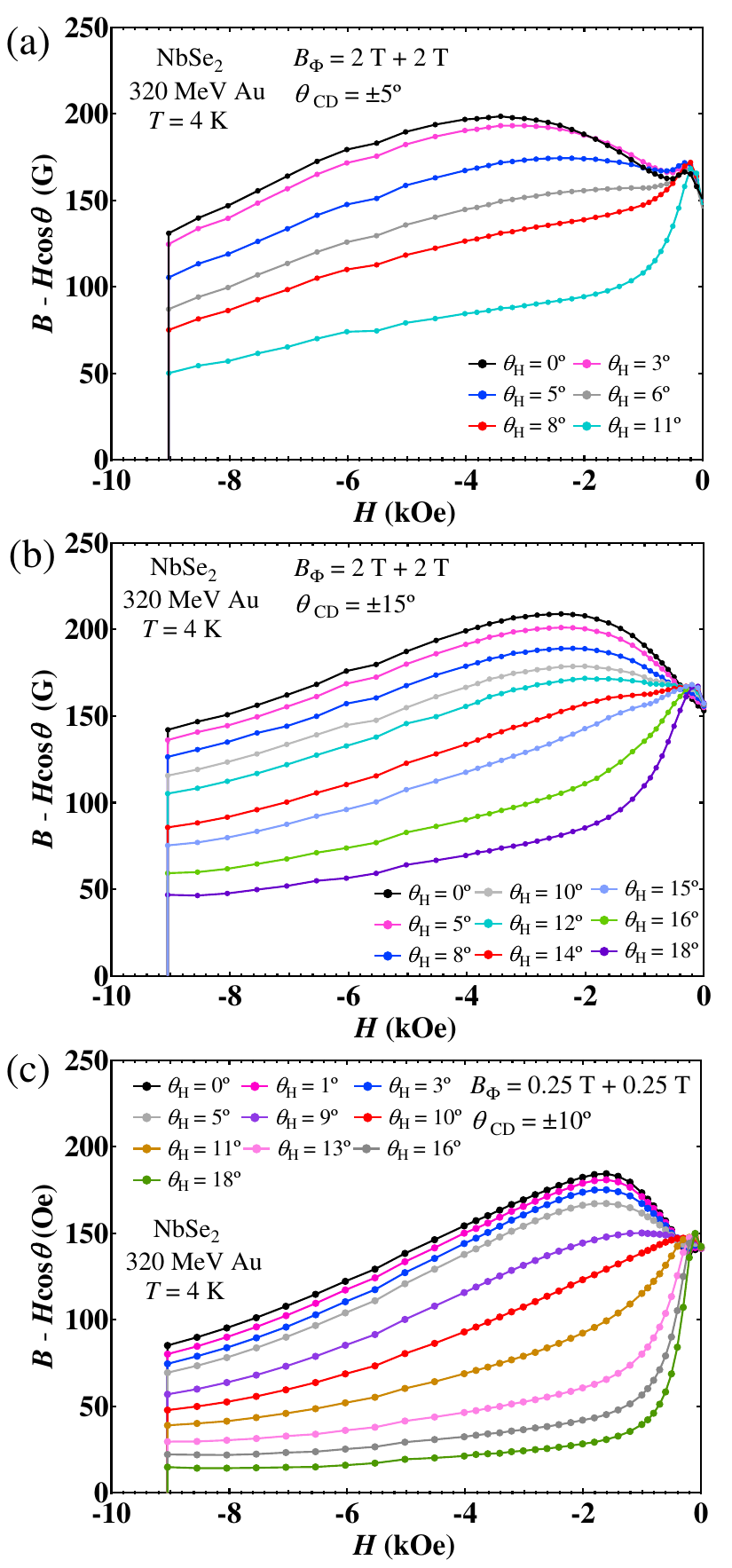}
\caption{\label{fig:13}Magnetic field dependence of \emph B - \emph Hcos$\theta_{\mathrm{}}$ at \emph T = 4 K at various $\theta_{\mathrm{\emph H}}$, for NbSe${\rm {_2}}$ single crystals with splayed CDs introduced by 320 MeV Au irradiation with (a) \emph B{${_\Phi}$} = 2 T + 2 T, $\theta_{\mathrm{CD}}$ = ±5°, (b) \emph B{${_\Phi}$} = 2 T + 2 T, $\theta_{\mathrm{CD}}$ = ±15°, and (c) \emph B{${_\Phi}$} = 0.25 T + 0.25 T, $\theta_{\mathrm{CD}}$ = ±10°. Here, magnetic induction (\emph B) is measured by a miniature Hall probe.}
\end{figure}

To further investigate the origin of the anomalous peak effect, we studied the effects of the angle dependence of the applied field $\theta_{\mathrm{\emph H}}$ on NbSe${\rm {_2}}$ with splayed CDs introduced by 320 MeV Au irradiation. The local induction parallel to the \emph c-axis generated by the superconducting current in the sample, subtracted by the component of the external field along the \emph c-axis, \emph B - \emph Hcos$\theta$, for three NbSe${\rm {_2}}$ samples with different \emph B{${_\Phi}$} and $\left| \theta_{\mathrm{CD}} \right|$ under various $\theta_{\mathrm{\emph H}}$ are shown in Fig. \ref{fig:13}, (a) \emph B{${_\Phi}$} = 2 T + 2 T and $\left| \theta_{\mathrm{CD}} \right|$ = 5°, (b) \emph B{${_\Phi}$} = 2 T + 2 T and $\left| \theta_{\mathrm{CD}} \right|$ = 15°, and (c) \emph B{${_\Phi}$} = 0.25 T + 0.25 T and $\left| \theta_{\mathrm{CD}} \right|$ = 10°. The shapes of \emph B - \emph Hcos$\theta_{\mathrm{}}$ vs.\emph H curves and values of \emph H${\rm {_p}}$ obtained by using SQUID magnetometer agree reasonably well with those measured by the Hall probe at $\theta_{\mathrm{\emph H}}$ = 0°. As $\theta_{\mathrm{\emph H}}$ is increased, the anomalous peak effect is gradually weakened and strongly suppressed above certain $\theta_{\mathrm{\emph H}}$, which is close to $\left| \theta_{\mathrm{CD}} \right|$ for all of the three measured samples. Actually, a similar phenomenon had also been observed in iron-based superconductors with splayed CDs \cite{park2018field}. In Ba$_{1-x}$K$_x$Fe$_2$As$_2$ introduced with splayed CDs (\emph B{${_\Phi}$} = 4 T + 4 T, $\left| \theta_{\mathrm{CD}} \right|$ = 15°), the anomalous peak effect was also strongly suppressed when the magnetic field was applied parallel to one of the splay CDs ($\theta_{\mathrm{\emph H}}$ = $\theta_{\mathrm{CD}}$ = 15°). These experimental results suggest that suppression of anomalous peak effect at $\theta_{\mathrm{\emph H}}$ = $\theta_{\mathrm{CD}}$ is a general feature in superconductors with symmetric splayed CDs.

The fact that the anomalous peak effect is observed in samples with splayed CDs, but not in a sample with CDs parallel to the \emph c-axis indicates that its origin should be related to elements that are present only in the former. Crossing points of two kinds of CDs is one of important elements only present in systems with splayed CDs. By applying the magnetic field along the average direction of two kinds of CDs (one along +$\theta_{\mathrm{CD}}$ and another -$\theta_{\mathrm{CD}}$), half of vortices are trapped by CDs along +$\theta_{\mathrm{CD}}$ and the rest half are trapped by CDs along -$\theta_{\mathrm{CD}}$ as long as the splay angle is not too large. In such a situation, the presence of vortex along +$\theta_{\mathrm{CD}}$ near the crossing point suppresses the motion of vortex along -$\theta_{\mathrm{CD}}$ near the same point since forced entanglement of these two kinds of vortices are energetically unfavorable. By tilting the magnetic field away from the average direction of two kinds of splayed CDs, population of one kind of vortices increases and make the situation similar to the case of parallel CDs along the \emph c-axis for \emph H parallel to CDs, where anomalous peak effect is not observed. One significant complication in this interpretation is that in the case of tilted CDs for magnetic field parallel to the CDs, similar anomalous peak effect has been observed \cite{civale1991vortex,eley2018glassy}. In a system with tilted CDs for \emph H parallel to CDs, dynamics of vortex kinks near the surface of the sample is important as suggested in Ref. \cite{schuster1996direct}. Whether anomalous peak effects observed in superconductors with symmetric splayed CDs and that observed in superconductors with tilted CDs originate from the same origin or not is not clear at the present stage. Some studies had reported similar features of peak effects in samples with tilted CDs \cite{li2021peak} and splayed CDs \cite{tamegai2023anomalous}. For both of these two peak effects, as the density of CDs increases, the \emph H${\rm {_p}}$ shifts from low fields to high fields. However, the entangled vortex state is formed in superconductors with splayed CDs when the field is applied along the average direction of splayed CDs, which may have some effects on the formation of the anomalous peak effect. Studies on superconductors with asymmetric splayed CDs with different \emph B{${_\Phi}$} for the two directions may be important to further understand the anomalous peak effect. By increasing \emph B{${_\Phi}$} of one direction from zero to the value of another direction of CDs, the CD system can be gradually transformed from tilted CDs to splayed CDs. Such a study may shed light on the relationship between anomalous peak effect induced by tilted CDs and by splayed CDs, providing valuable information on the mechanism of the anomalous peak effect.

\section{\label{sec:level2}Summary}
Three kinds of peak effects were observed in NbSe${\rm {_2}}$ single crystals irradiated by 3 MeV proton, 320 MeV Au, and 800 MeV Xe irradiations. Pronounced peak effects originated from O-DO transition of vortices were observed in 3 MeV proton irradiated crystals. With increasing proton dose, the \emph H${\rm {_p}}$ shifts from high fields to low fields. Similar O-DO transition-induced peak effects were also observed in NbSe${\rm {_2}}$ with splayed CDs introduced by 320 MeV Au and 800 MeV Xe irradiations. The experimental results demonstrate that in NbSe${\rm {_2}}$ with splayed CDs, as $\left| \theta_{\mathrm{CD}} \right|$ increases, \emph H${\rm {_p}}$ of the self-field peak effect gradually shifts from high fields to low fields. The simulation results agree well with experimental observations. Detailed analyses based on the simulation results of field dependence of magnetic moment reveals the formation of peculiar non-uniform \emph J${\rm {_c}}$ flow as $\left| \theta_{\mathrm{CD}} \right|$ increases, which plays a vital role in inducing the self-field peak effect in superconductors with splayed CDs. It is also found that two kinds of peak effects can coexist in certain conditions. Coexistence of O-DO transition-induced peak effect and self-field peak effect was confirmed for both NbSe${\rm {_2}}$ irradiated by 800 MeV Xe and 320 MeV Au. In addition, it is found that the anomalous peak effect is strongly suppressed in NbSe${\rm {_2}}$ with symmetric splayed CDs when the magnetic field is applied parallel to one of the splay CDs. This observation is consistent with that in iron-based superconductors, and raise an important question whether anomalous peak effects observed in superconductors with symmetric splayed CDs and tilted CDs share a common origin or not.

\nocite{*}

\section*{References}
\bibliography{iopart-num}

\end{document}